\documentclass[10pt,twocolumn]{article}

\usepackage{amsmath,amssymb,amsfonts}
\usepackage{amsthm}
\usepackage{bm}
\usepackage[hidelinks]{hyperref}
\usepackage[a4paper,margin=1.7cm,columnsep=0.65cm]{geometry}
\usepackage{cite}

\newtheorem{theorem}{Theorem}
\newtheorem{proposition}{Proposition}

\newcommand{\cH}{\mathcal H}
\newcommand{\cA}{\mathcal A}
\newcommand{\cC}{\mathcal C}
\newcommand{\cL}{\mathcal L}
\newcommand{\Tr}{\operatorname{Tr}}
\newcommand{\id}{\mathbb I}
\newcommand{\EB}{\mathrm{EB}}
\newcommand{\JM}{\mathrm{JM}}
\newcommand{\CPTP}{\mathrm{CPTP}}
\newcommand{\dd}{\mathrm d}
\newcommand{\dav}{d_{\mathrm{av}}}

\allowdisplaybreaks
\title{Minimum Experimental Information for Certifying Bell Nonlocality of Gravity}
\author{Jianqi Sheng\\
\normalsize Department of Physics, City University of Hong Kong, Hong Kong\\
\normalsize \texttt{sheng.jq@cityu.edu.hk}}
\date{}

\begin{document}

\twocolumn[
\begin{@twocolumnfalse}
\maketitle
\vspace{-1.5em}
\begin{abstract}
A Bell violation by material probes certifies nonlocal behavior, but it does not
identify which internal subsystem carried or was measured to produce that
behavior.  We determine the additional information required to attribute such a
violation to a gravitational field.  For any setting-independent field-to-probe
channel completed before the Bell settings are chosen, an
entanglement-breaking channel makes all induced field measurements jointly
measurable.  A static Newtonian phase pickup belongs to this class, even with
arbitrary adaptive probe control.  Applied to a recent four-mass proposal, the
claimed field-only measurements obey the stronger CHSH bound \(\sqrt2\), whereas
the Bell-active operation is the source-to-probe instrument that includes the
coherent source herald.  At the published parameters, the observed flagged
behavior is at least \(0.1630\) from the claimed field-detector family, exceeding
its \(0.0898\) distance from Bell locality.  We then incorporate a common field
slice, gravitational constraints, dressing, detector backreaction, and coherent
references into a robust certification theorem.  The result isolates the
minimum missing information: field-only factorization and a local measurement-
incompatibility witness on each wing.  More precise phase data cannot replace
either requirement.
\end{abstract}
\vspace{0.75em}
\end{@twocolumnfalse}
]

\section{Introduction}

Bell experiments certify correlations without identifying the microscopic
carrier inside each device.  This separation is an advantage when the objective
is device-independent nonlocality.  It becomes a limitation when the conclusion
names an inaccessible subsystem, such as the gravitational field.  A final Bell
violation can be compatible with a quantum gravitational carrier, a different
quantum bypass, or a local sequential instrument acting on pre-entangled matter.
These realizations can induce the same observable process on the material ports.

This issue is immediate in proposals that begin with two entangled source masses
and couple each source to a nearby probe.  The probes can inherit the initial
cross-laboratory entanglement through two local interfaces.  A Bell violation by
the probes then certifies the final behavior and, under a clean resource ledger,
nonclassical local transduction.  It does not by itself locate an entangled state
on two gravitational field algebras at an intermediate time.

Das \emph{et al.}\ recently proposed a four-mass experiment intended to close
this gap \cite{Das2026}.  Their protocol first prepares two distant source spins
in a Bell state, maps the spins to source-position superpositions, and lets two
nearby probes acquire local Newtonian phases.  The source paths are then closed
and their spins are measured in the \(X\) basis (Step 5).  The probe paths are
closed afterward, and CHSH settings are chosen on the probe spins (Step 6).
Steps 4 and 6 are interpreted as measurements of the source gravitational
fields, while Step 5 is described as a local assisting herald.

We analyze the operational content of this interpretation.  The central result
is not a general prohibition on Bell tests of gravity.  It is a boundary between
three distinct claims: Bell nonlocality of the final probe records, quantum
transduction through a gravity-labelled interface, and Bell nonlocality of a
specified field state.  The boundary is fixed by the detector effects, their
time slice, and their algebraic support.

\section{Operational Claims and the Detector Criterion}

\paragraph{Certification ladder.}
Let \(P_0\) denote an event-ready Bell violation by the registered probe
outcomes.  Let \(P_1\) denote simultaneous non-entanglement-breaking resource in
the two complete local source-to-probe interfaces.  Let \(P_2\) attribute this
resource to gravity after excluding quantum bypasses.  Let \(P_3\) assert an
entangled state on two specified gravitational field algebras on one common
slice.  Finally, let \(P_4\) denote Bell-nonlocal statistics generated by
field-only measurements on those algebras.

The implications are not automatic.  Under Bell-valid event-ready selection and
a clean initial-resource ledger, \(P_0\) lower-bounds a joint version of \(P_1\).
The step \(P_1\to P_2\) requires a carrier-sensitive quantum cut.  A coherent
reset bus shows that \(P_2\not\Rightarrow P_3\), because a mediator can return to
its initial state after transferring entanglement.  An entangled state does not
violate a Bell inequality under jointly measurable local effects, so
\(P_3\not\Rightarrow P_4\).  Only \(P_4\), relative to specified field algebras
and Bell assumptions, certifies \(P_3\).

The correct joint resource for \(P_1\) is also not the pair of averaged marginal
channels.  Shared classical histories can make each average marginal EB while a
correlated joint channel preserves a Bell state.  The relevant free set is
\begin{equation}
 \mathsf F_{\wedge}=\overline{\operatorname{conv}}
 \big[(\EB_A\otimes\CPTP_B)\cup
 (\CPTP_A\otimes\EB_B)\big].
 \label{eq:andfree}
\end{equation}
For a shared-history decomposition with conditional local channels and
normalized diamond distances \(r_j(\lambda)\) to \(\EB\), event-ready Bell data
give
\begin{equation}
 \mathbb E_\lambda\min\{r_A(\lambda),r_B(\lambda)\}
 \ge \dav(p,\cL_{\rm ER}).
 \label{eq:andbound}
\end{equation}
This is a process statement.  It does not identify a field state.

\paragraph{Fixed-preprocessing detector theorem.}
Consider one wing.  A setting-independent interaction is completed before the
Bell setting \(x\) is selected and defines a field-to-probe channel
\begin{equation}
 \Lambda:\mathsf T(\cH_G)\longrightarrow\mathsf T(\cH_P).
 \label{eq:channel}
\end{equation}
The later probe POVM \(\{F^P_{a|x}\}_a\) induces the field effect
\begin{equation}
 E^G_{a|x}=\Lambda^*(F^P_{a|x}).
 \label{eq:pullback}
\end{equation}

\begin{theorem}[Fixed-preprocessing detector]
If \(\Lambda\) is entanglement breaking, all POVMs induced by arbitrary
downstream probe readouts are jointly measurable.  Consequently, if either wing
has EB preprocessing, no bipartite field state can violate a Bell inequality in
this delayed-choice architecture.
\end{theorem}

Every EB channel has a measure-and-prepare representation
\cite{Horodecki2003}
\begin{equation}
 \Lambda(\rho)=\int\Tr[M(\dd\lambda)\rho]\,\sigma_\lambda.
 \label{eq:holevo}
\end{equation}
Its adjoint maps each pointer effect to
\begin{equation}
 E_{a|x}=\int M(\dd\lambda)
 \Tr(\sigma_\lambda F_{a|x}).
 \label{eq:parent}
\end{equation}
Thus every setting is a classical postprocessing of the same parent POVM
\(M\).  Joint measurability precludes Bell nonlocality independently of the
state and of the other wing's measurements \cite{Wolf2009}.  Non-EB
preprocessing is necessary here, but it is not sufficient.  Some non-EB
channels still break measurement incompatibility \cite{Heinosaari2015}.  The
exact local requirement for the selected settings is
\begin{equation}
 \{\Lambda^*(F_{a|0}),\Lambda^*(F_{a|1})\}_a\notin\JM.
 \label{eq:exactlocal}
\end{equation}

The temporal qualification matters.  A setting-dependent destructive
measurement may realize incompatible field POVMs even though each individual
quantum-to-classical readout is EB.  The theorem applies when all setting
dependence is downstream of one fixed channel, as in Steps 4 and 6 of
Ref.~\cite{Das2026}.

\section{No-Go Result for Static Newtonian Phase Detection}
Let the probe couple to a field configuration observable \(\Phi\) through
\begin{equation}
 U_{GP}=\int P_\Phi(\dd\lambda)\otimes U^P_\lambda,
 \label{eq:qnd}
\end{equation}
where \(P_\Phi\) is its spectral measure.  For an arbitrary probe state
\(\sigma_P\), Eq.~\eqref{eq:pullback} becomes
\begin{equation}
 E^G_{a|x}=\int P_\Phi(\dd\lambda)
 \Tr(\sigma_PU_\lambda^\dagger F^P_{a|x}U_\lambda).
 \label{eq:abelianeffects}
\end{equation}
The corresponding channel is
\begin{equation}
 \Lambda_{G\to P}(\rho_G)=\int\Tr[P_\Phi(\dd\lambda)\rho_G]
 U_\lambda\sigma_PU_\lambda^\dagger.
 \label{eq:ebphase}
\end{equation}
It is explicitly EB.  The proof extends to a complete commuting family of
configuration observables.

The conclusion also survives arbitrary probe-side elaboration.  Suppose every
field--probe interaction in a finite adaptive network is decomposable over the
same spectral measure,
\begin{equation}
 U_k=\int P(\dd\lambda)\otimes U_{k,\lambda}.
 \label{eq:adaptive}
\end{equation}
Probe ancillas, intermediate probe measurements, feed-forward, and
setting-dependent probe channels can be dilated coherently.  The complete
network remains decomposable over \(P\), and every final effect has the form
\begin{equation}
 E_{a|x}=\int P(\dd\lambda)p(a|x,\lambda).
 \label{eq:adaptiveeffect}
\end{equation}
Repeated phase acquisition cannot supply the missing incompatibility.  An escape
requires a noncommuting field operator, non-QND field dynamics, or a setting that
directly selects a different field interaction.

This separates phase sensitivity from quantum readout capacity.  For
\(U=\exp(-ig\Phi\otimes Z_P)\) and \(|+x\rangle_P\), a field eigenvalue produces
\(|\psi_\lambda\rangle=e^{-ig\lambda Z}|+x\rangle\).  This family has quantum
Fisher information \(4g^2\) and can distinguish different eigenvalues, while
the channel in Eq.~\eqref{eq:ebphase} remains EB and has zero quantum capacity.
A detector can measure a classical field parameter accurately without
transferring field coherence.

Detector-based measurement theory in QFT indeed associates a field POVM with a
localized detector coupling \cite{PoloGomez2022}.  For detector input
\(|\psi\rangle\), interaction \(U\), and detector result \(|s\rangle\), the
field Kraus operator is \(M_{s,\psi}=\langle s|U|\psi\rangle\), with effect
\(M_{s,\psi}^\dagger M_{s,\psi}\).  This construction proves that a POVM exists.
It does not prove that two detector bases induce incompatible POVMs.

\section{Application to the Four-Mass Protocol}
For the ideal phase pickup
\begin{equation}
 U=\exp(-i\Phi\otimes Z_P),\qquad
 \sigma_P=|+x\rangle\langle+x|,
 \label{eq:idealphase}
\end{equation}
the induced dichotomic observables are
\begin{equation}
 X_P\mapsto\cos(2\Phi),\qquad
 Y_P\mapsto\sin(2\Phi),\qquad
 Z_P\mapsto0.
 \label{eq:pullbackpauli}
\end{equation}
The settings in Ref.~\cite{Das2026} are
\begin{equation}
 A_0=X,\quad A_1=-Z,\qquad
 B_{0,1}=\frac{X\pm Z}{\sqrt2}.
 \label{eq:papersettings}
\end{equation}
They induce \(A_0=\cos(2\Phi_A)\), \(A_1=0\), and
\(B_0=B_1=\cos(2\Phi_B)/\sqrt2\).  Hence every field state obeys
\begin{equation}
 |S_{\rm field}|=\sqrt2\,
 |\langle\cos(2\Phi_A)\cos(2\Phi_B)\rangle|
 \le\sqrt2.
 \label{eq:sqrt2}
\end{equation}
The field-only setting pair is not a Bell analyzer.  The larger CHSH values in
the full protocol must come from an operation absent from
Eqs.~\eqref{eq:idealphase}--\eqref{eq:papersettings}.

\paragraph{The Bell-active source instrument.}
After the two local controlled phase interactions, the four-matter-qubit
calculation has the normal form
\begin{equation}
 |\Omega\rangle=\frac{1}{\sqrt2}
 \big(|00\rangle_S|u_0v_0\rangle_P
 +|11\rangle_S|u_1v_1\rangle_P\big).
 \label{eq:normalform}
\end{equation}
Before Step 5, the reduced probe state is the separable mixture
\begin{equation}
 \rho_P=\frac12\sum_{j=0}^1|u_jv_j\rangle\langle u_jv_j|.
 \label{eq:probeseparable}
\end{equation}
Local source measurements in the \(X\) basis, with product outcome sign
\(\eta=\pm1\), prepare
\begin{equation}
 |\psi_\eta\rangle_P\propto
 |u_0v_0\rangle+\eta|u_1v_1\rangle.
 \label{eq:conditionalprobe}
\end{equation}
Step 5 is therefore an entanglement-of-assistance operation.  It does not
increase entanglement across the full laboratory cut
\((S_AP_A):(S_BP_B)\), but it can transfer the pre-existing source ebit to the
different cut \(P_A:P_B\).  Locality across the laboratories does not imply
locality across a source--field cut.

Let \(s_A=\langle u_0|u_1\rangle\) and
\(s_B=\langle v_0|v_1\rangle\).  The probability of a particular source
outcome pair and the conditional probe concurrence are
\begin{align}
 p_\eta&=\frac{1+\eta\operatorname{Re}(s_As_B)}4,
 \label{eq:peta}\\
 C_\eta&=\frac{\sqrt{(1-|s_A|^2)(1-|s_B|^2)}}
 {1+\eta\operatorname{Re}(s_As_B)}.
 \label{eq:ceta}
\end{align}
For the symmetric protocol, \(s_A=s_B=\cos\theta\).  Combining the two outcome
pairs of each sign gives
\begin{align}
 P_+&=\frac{1+\cos^2\theta}{2},&
 C_+&=\frac{\sin^2\theta}{1+\cos^2\theta},
 \nonumber\\
 P_-&=\frac{\sin^2\theta}{2},& C_-&=1.
 \label{eq:branches}
\end{align}
With the fixed Step-6 axes, the two conditional CHSH values are
\begin{equation}
 S_+=\frac{2\sqrt2}{1+\cos^2\theta},\qquad
 S_-=2\sqrt2.
 \label{eq:conditionalS}
\end{equation}
The full pre-setting-flag functional is consequently
\begin{equation}
 W=\sum_h p_hS_h=\sqrt2(1+\sin^2\theta).
 \label{eq:W}
\end{equation}
The probability-weighted concurrence is
\(C_{\rm av}=\sin^2\theta\), so
\begin{equation}
 W-\sqrt2=\sqrt2 C_{\rm av}.
 \label{eq:mechanismidentity}
\end{equation}
For this protocol, the entire excess over the Step-4-plus-Step-6 detector bound
tracks the source entanglement concentrated into the probes by Step 5.

The distinction is also exact at channel level.  Including the source herald,
the local source-to-probe flagged channel has Kraus operators, up to local basis
phases,
\begin{equation}
 K_+=\frac{\id+e^{i\theta}X}{2},\qquad K_-=K_+Z,
 \label{eq:kraus}
\end{equation}
and satisfies
\begin{equation}
 \operatorname{dist}_\diamond(\widehat{\mathcal T}_\theta,\EB)
 =\frac{|\sin\theta|}{2}.
 \label{eq:exactEBdistance}
\end{equation}
Here
\(\operatorname{dist}_\diamond(\mathcal N,\EB)
=\inf_{\mathcal E\in\EB}\|\mathcal N-\mathcal E\|_\diamond/2\)
is the normalized diamond distance.
This resource belongs to the flagged source-to-probe instrument, not to the
field-to-probe channel in Eq.~\eqref{eq:ebphase}.

Suppose instead that Step 5 prepared a field state on a common slice through an
instrument \(\{\Gamma_z:S\to G\}\), after which the static detector
\(\Lambda_{G\to P}\) acted.  The complete flagged map would be
\begin{equation}
 \widehat{\mathcal T}_{\rm fact}(\rho_S)=
 \sum_z\Lambda(\Gamma_z(\rho_S))\otimes|z\rangle\langle z|.
 \label{eq:factorizedmap}
\end{equation}
Because \(\Lambda\) is EB, every map of the form
Eq.~\eqref{eq:factorizedmap} is EB.  Equation~\eqref{eq:exactEBdistance}
therefore implies
\begin{equation}
 \inf_{\{\Gamma_z\}}\frac12
 \|\widehat{\mathcal T}_\theta-
 \widehat{\mathcal T}_{\rm fact}\|_\diamond
 \ge\frac{|\sin\theta|}{2}.
 \label{eq:factorizationgap}
\end{equation}
At the published value \(\theta=1.85369756598\), the lower bound is
\(0.48012\).  It reaches \(0.5\) at \(\theta=\pi/2\).  The matter calculation
cannot be represented as a common-slice field preparation followed by the
claimed static field detector.

\section{Field Attribution on a Common Slice}
An assisted Bell test of a field state requires a slice \(\Sigma\) on which the
heralded field state exists before the field measurements.  Consider a local
branch-sensitive interaction
\begin{equation}
 U_{SP}=|0\rangle\langle0|\otimes U_0
 +|1\rangle\langle1|\otimes U_1.
 \label{eq:controlled}
\end{equation}
A coherent source flip can be commuted through this interaction, up to source
phases, if and only if
\begin{equation}
 U_1^\dagger U_0\propto\id_P.
 \label{eq:deferralcondition}
\end{equation}
This is precisely the condition that the probe acquires no branch-dependent
quantum information.  If \(\chi\) is the entangling phase invariant, the
phase-optimized operator-norm obstruction is
\begin{equation}
 \epsilon_{\rm def}=2\sin\left[
 \frac{d_{2\pi}(\chi,0)}4\right].
 \label{eq:deferral}
\end{equation}
For the four-mass gate, \(\chi=-2\theta\).  The published value gives
\(\epsilon_{\rm def}=1.20071417641\), while the Bell-optimal point gives
\(\sqrt2\).  The useful signal and the ability to advance Step 5 are mutually
exclusive in this controlled model.

This produces a fork.  If Step 5 is a source-only herald that can be moved
before \(\Sigma\), the later static field detector is jointly measurable and
cannot violate Bell.  If Step 5 cannot be moved, it is part of the Bell-active
sequential instrument.  The result remains a valid late-probe Bell test, but it
is not a Bell test of a field state under field-only POVMs.

\paragraph{Gravitational constraints and dressing.}
The obstruction is not resolved by denying that a Newtonian field can have a
quantum state.  Linearized gravity admits source-correlated field wave
functionals \cite{Chen2022}, and the longitudinal/radiative distinction alone
does not settle the carrier question \cite{Christodoulou2022,Danielson2022}.
The operational issue is the physical algebra and its allowed interventions.

For a branch-dependent source density,
\begin{equation}
 \rho_S(x)=\rho_0(x)|0\rangle\langle0|
 +\rho_1(x)|1\rangle\langle1|,
 \end{equation}
the bare source flip fails to commute with the constraint
\(C(x)=C_G(x)-\rho_S(x)\).  A physical coherent branch operation must carry a
field dressing,
\begin{equation}
 \widetilde X=|1\rangle\langle0|\otimes D_{10}
 +|0\rangle\langle1|\otimes D_{01}.
 \label{eq:dressedX}
\end{equation}
This is consistent with the general need for gravitationally dressed
observables \cite{Donnelly2016} and with recent gauge-constrained analyses of
field-mediated entanglement \cite{Spalvieri2026}.  The final bare spin readout
may occur after the source paths close.  Pulling the complete Step-5 instrument
back to the branch-distinct slice nevertheless includes the field translation
associated with recombination.

A global source--field tensor product is not required.  In a physical
shared-center decomposition,
\begin{equation}
 \cH_{\rm phys}=\int^\oplus\dd q\,
 \cH_S(q)\otimes\cH_G(q),
 \label{eq:directintegral}
\end{equation}
a field-only effect has the form
\begin{equation}
 E_a^G=\int^\oplus\dd q\,
 \id_{S(q)}\otimes E_a^G(q).
 \label{eq:fieldeffect}
\end{equation}
The common center is classical shared information.  A field Bell violation
must arise from noncentral, incompatible effects inside at least one fixed
sector.  If the quasi-static Newtonian fiber is one dimensional after boundary
data are fixed, that sector cannot support a direct field Bell analyzer.  Gauge
constraints also make subsystem independence nontrivial
\cite{BoulleFranzmann2026}; Eq.~\eqref{eq:fieldeffect} states the operational
factorization that the experiment must validate.

A coherent reference can enable effects that connect otherwise inaccessible
sectors.  It then becomes a resource.  The two wing references must be initially
separable, their asymmetry or edge charge must be calibrated, and their induced
effects must remain identity on the entangled source multiplicities.

The Newtonian phase is also reciprocal.  With source and probe densities, the
constraint solution has energy
\begin{equation}
\begin{split}
 E[\rho_S+\rho_P]={}&-\frac{1}{2}
 \langle\rho_S,K^{-1}\rho_S\rangle
 -\frac{1}{2}\langle\rho_P,K^{-1}\rho_P\rangle\\
 &-\langle\rho_S,K^{-1}\rho_P\rangle.
\end{split}
\label{eq:reciprocity}
\end{equation}
The measured phase comes from the symmetric cross term.  Calling the probe a
test mass does not select a one-way measurement of the source field.  A field
interpretation needs an independently specified field algebra and a bound on
the disturbance caused by inserting the probe.

\section{Robust Certification}
Let \(\delta_V\) bound Bell-validity errors, and let \(\delta_\Sigma\) bound the
change induced by replacing the actual timing with a herald-before-measurement
common-slice realization.  Let \(\delta_F\) bound the replacement of the
pulled-back effects by the field-only form in Eq.~\eqref{eq:fieldeffect},
including localization error.  Finally, let \(\epsilon_{\rm ref}\) and
\(\delta_{\rm back}\) bound, respectively, the removal of illicit cross-wing
resource in detector references and the removal of probe backreaction on the
declared field state.  Every quantity is an upper bound on induced average
total variation distance, or on a strategy norm that contracts to it.

\begin{proposition}[Robust field attribution]
If these five replacements produce a behavior \(p_{\rm field}\) generated by
field-only measurements on the declared common-slice algebras, then
\begin{equation}
 \begin{split}
 \dav(p_{\rm field},\cL_{\rm ER})\ge
 \Big[&\dav(p_{\rm obs},\cL_{\rm ER})-\delta_V-\delta_\Sigma\\
 &-\delta_F-\epsilon_{\rm ref}-\delta_{\rm back}\Big]_+ .
 \end{split}
 \label{eq:robust}
\end{equation}
\end{proposition}
A positive right-hand side is sufficient for \(P_4\) relative to the specified
field algebras.  Each correction has a different operational role, so none can
be inferred from the endpoint CHSH value.

The published behavior makes the missing detector term quantitative.  For
uniform settings, a CHSH difference is bounded by eight times the average total
variation distance.  Since the specified field detector satisfies
Eq.~\eqref{eq:sqrt2},
\begin{equation}
 \dav(p_h,\mathsf B_{\rm det})
 \ge\frac{(S_h-\sqrt2)_+}{8}.
 \label{eq:conditionaldistance}
\end{equation}
At \(\theta=1.85369756598\), the equal-sign branches have
\(S_h=2.62396595145\), giving \(0.15121904863\).  The crossed branches give
\(0.17677669530\).  For the full flagged functional,
\begin{equation}
 W=2.71823066275,
 \end{equation}
and
\begin{align}
 \dav(p_{\rm full},\mathsf B_{\rm det})
 &\ge\frac{W-\sqrt2}{8}=0.16300213755,
 \label{eq:detdist}\\
 \dav(p_{\rm full},\cL_{\rm ER})
 &\ge\frac{W-2}{8}=0.08977883284.
 \label{eq:localdist}
\end{align}
Any strategy norm used for \(\delta_F\) must dominate the induced behavior
distance.  Thus \(\delta_F\ge0.16300213755\) for the claimed detector target.
Even if every other correction in Eq.~\eqref{eq:robust} vanished, the resulting
field certificate would be zero.

The minimum missing local datum can be stated without full tomography.  For two
binary field POVMs, joint measurability is the feasibility problem
\begin{align}
 G_{\alpha\beta}&\ge0,\qquad
 \sum_{\alpha,\beta}G_{\alpha\beta}=\id,
 \nonumber\\
 E_{a|0}&=\sum_\beta G_{a\beta},\qquad
 E_{b|1}=\sum_\alpha G_{\alpha b}.
 \label{eq:jmfeasibility}
\end{align}
One dual witness separating the calibrated effects from this convex set is
sufficient.  For unbiased qubit effects with Bloch vectors \(\bm a_0,\bm a_1\),
the condition is
\begin{equation}
 \|\bm a_0+\bm a_1\|+\|\bm a_0-\bm a_1\|>2.
 \label{eq:blochjm}
\end{equation}
The witness must refer to effects satisfying Eq.~\eqref{eq:fieldeffect}; a
source-assisted calibration would only recertify the flagged source-to-probe
instrument.

This coherence-sensitive information is necessary.  Calibration on branch-
diagonal field inputs determines a qubit detector only on
\(\operatorname{span}\{\id,Z\}\).  The identity and complete dephasing channels
agree on this operator system, although one is coherent and the other EB.  No
number of additional diagonal phase measurements distinguishes them.  At least
one off-diagonal channel or incompatibility functional is required.

\paragraph{A nonempty positive boundary.}
The no-go applies to the static QND detector, not to every field detector.  Let a
fixed physical sector contain a field code
\(\cC_G=\operatorname{span}\{|0_L\rangle,|1_L\rangle\}\), and suppose a local
isometry implements
\begin{equation}
 V|j_L\rangle_G|0\rangle_P=|r\rangle_G|j\rangle_P.
 \label{eq:resettransfer}
\end{equation}
It resets the residual field record and transfers both field coherences to the
probe.  Probe \(X\) and \(Z\) measurements then pull back to incompatible
\(X_L\) and \(Z_L\) field effects.

A fixed-charge dual-rail realization uses two degenerate field modes and two
detector modes,
\begin{equation}
 |0_L\rangle=g_0^\dagger|\mathrm{vac}\rangle,\qquad
 |1_L\rangle=g_1^\dagger|\mathrm{vac}\rangle,
\end{equation}
with the number-conserving interaction
\begin{equation}
 H_{\rm swap}=\hbar\kappa\sum_{j=0,1}
 (g_j^\dagger p_j+p_j^\dagger g_j).
 \label{eq:swap}
\end{equation}
At time \(\pi/(2\kappa)\), the field code is swapped to the detector.  This
model proves that the sufficient boundary is nonempty.  For gravity, it would
require noncentral radiative or curvature modes and gravity-selective exchange.
The static Newtonian phase model implements neither ingredient.

\section{Conclusion}
A final Bell violation answers a behavioral question.  A claim about the
gravitational field additionally requires information that fixes the carrier,
time slice, and measurement algebra.  In a delayed-choice indirect detector,
the irreducible local requirement is an incompatibility witness for field-only
pulled-back effects.  The four-mass protocol supplies nonlocal late-probe
behavior and a non-EB source-to-probe flagged instrument.  Its static Newtonian
phase pickup supplies only compatible field effects, while the coherent source
herald provides the missing Bell-active operation.

The minimum information is therefore not another estimate of the gravitational
phase.  It is a common-slice field-factorization certificate, one local
field-measurement incompatibility witness per wing, and a closed ledger for
quantum bypasses, coherent references, and detector backreaction.  These data
are sufficient through Eq.~\eqref{eq:robust} and necessary because the endpoint
phase statistics leave EB and coherent detector extensions in the same
observational fiber.

\appendix

\section{Proof of the joint-resource bound}

Condition on a setting-independent event-ready flag, and let the resulting
normalized joint channel have the input-independent shared-history form
\begin{equation}
 \mathcal N=\int\mu(\dd\lambda)\,
 \mathcal N_A^\lambda\otimes\mathcal N_B^\lambda .
\end{equation}
For any \(\varepsilon>0\), choose an EB channel
\(\mathcal E_j^\lambda\) satisfying
\begin{equation}
 \frac12\|\mathcal N_j^\lambda-\mathcal E_j^\lambda\|_\diamond
 \le r_j(\lambda)+\varepsilon .
\end{equation}
For each \(\lambda\), replace the wing with smaller \(r_j(\lambda)\), and call
the averaged channel \(\mathcal M_\varepsilon\).  It belongs to
\(\mathsf F_\wedge\).  Convexity and stability of the diamond norm under a
tensor factor that is CPTP give
\begin{equation}
 \frac12\|\mathcal N-\mathcal M_\varepsilon\|_\diamond
 \le\int\mu(\dd\lambda)\,
 \big[\min_j r_j(\lambda)+\varepsilon\big].
\end{equation}
A fixed Bell preparation-and-measurement tester contracts normalized diamond
distance to average total variation distance.  Every behavior obtained from
\(\mathcal M_\varepsilon\) is event-ready Bell local because at least one wing
is EB in every classical branch.  Taking the infimum over local behaviors and
then \(\varepsilon\to0\) proves Eq.~\eqref{eq:andbound}.  If the herald is not
setting independent, or if the classical history is correlated with the
unknown channel input, the bound requires an explicit enlarged strategy and
does not follow from the marginal channels alone.

\section{Proof of the adaptive abelian closure}

Let \(\cA\) be an abelian von Neumann algebra with joint spectral measure
\(P(\dd\lambda)\).  Dilate every probe operation and intermediate probe
measurement to an isometry on the probe plus local record systems.  By
assumption, each field-dependent isometry is decomposable over \(P\).  Products,
direct sums controlled by probe records, and setting-dependent probe-only
operations preserve this form.  The complete isometry for setting \(x\) is
therefore
\begin{equation}
 V_x=\int P(\dd\lambda)\otimes V_{x,\lambda}.
\end{equation}
Tracing the residual field gives
\begin{equation}
 \Lambda_x(\rho)=\int\Tr[P(\dd\lambda)\rho]\,
 \sigma_{x,\lambda},
\end{equation}
where \(\sigma_{x,\lambda}\) includes every probe ancilla and classical record.
This is measure-and-prepare.  A final effect produces
\begin{equation}
 E_{a|x}=\int P(\dd\lambda)p(a|x,\lambda).
\end{equation}
The same \(P\) is a parent POVM for all settings.  The bipartite LHV model is
obtained by measuring \(P_A\otimes P_B\) on the field state and using
\(p_A(a|x,\lambda)\) and \(p_B(b|y,\mu)\) as local response functions.

\section{Exact distance of the flagged phase instrument}

In the \(X\) basis, the singular values of \(K_+\) in
Eq.~\eqref{eq:kraus} are
\(c=|\cos(\theta/2)|\) and \(s=|\sin(\theta/2)|\); the other flag swaps the two
input basis vectors.  Delete the cross terms between the two rank-one
singular-vector maps in each flag.  The resulting flagged channel is EB and
trace preserving.

For an arbitrary input, including a reference, write the positive block matrix
in the \(X\) basis as
\(\bigl(\begin{smallmatrix}A&B\\B^\dagger&D\end{smallmatrix}\bigr)\).
Each flagged difference has trace norm \(2|cs|\|B\|_1\).  Positivity and unit
trace give \(\|B\|_1\le1/2\).  Summing the orthogonal flags yields a normalized
diamond distance no larger than \(|\sin\theta|/2\).

For the matching lower bound, apply the channel to one half of a normalized
maximally entangled state.  Each flag occurs with probability \(1/2\), and its
conditional pure Choi state has Schmidt coefficients \(c,s\).  Let
\(|\nu_h\rangle\) be the negative-eigenvalue eigenvector of the partial
transpose in flag \(h\), and set
\(W_h=(|\nu_h\rangle\langle\nu_h|)^{T_R}\).  Then
\(W_h\) is nonnegative on separable states,
\(\|2W_h\|_\infty=1\), and its expectation on the conditional Choi state is
\(-cs\).  Output dephasing in the flag basis fixes the target channel and maps
every EB comparison channel to another EB channel without increasing distance.
We may therefore restrict to block-diagonal separable comparison states.  The
dual operator \(H=2\bigoplus_hW_h\) has \(\|H\|_\infty=1\); its expectation on
the target Choi state is \(-2cs\) and on every comparison state is
nonnegative.  Trace-norm duality gives a normalized output distance at least
\(cs=|\sin\theta|/2\).  The diamond distance is no smaller than this
maximally-entangled-input distance, matching the upper bound and proving
Eq.~\eqref{eq:exactEBdistance}.

\section{Deferral obstruction}

For Eq.~\eqref{eq:controlled}, a source flip with an adjustable source phase
can commute through \(U_{SP}\) precisely when \(U_1=e^{i\alpha}U_0\).  Define
\begin{equation}
 \epsilon_{\rm def}=\inf_\alpha
 \|U_1-e^{i\alpha}U_0\|_{\rm op}.
\end{equation}
If the two eigenphases of \(U_1^\dagger U_0\) have shortest circular separation
\(d\), the optimal \(\alpha\) is the midpoint of that arc.  Elementary circle
geometry gives \(\epsilon_{\rm def}=2\sin(d/4)\).  The separation is the
distance modulo \(2\pi\) of the controlled-gate invariant
\(\chi=\phi_{00}+\phi_{11}-\phi_{01}-\phi_{10}\) from zero, yielding
Eq.~\eqref{eq:deferral}.

\section{Robust behavior distances}

For uniform binary settings, define
\begin{equation}
 \dav(p,q)=\frac18\sum_{x,y}\|p(\cdot|x,y)-q(\cdot|x,y)\|_1.
\end{equation}
The CHSH coefficients have magnitude one, so
\begin{equation}
 |S(p)-S(q)|\le8\dav(p,q).
\end{equation}
The same inequality holds for the full pre-setting-flag functional after a
fixed flag-dependent output relabeling.  Applying the detector bound
\(W\le\sqrt2\) gives Eq.~\eqref{eq:detdist}; applying the local bound \(W\le2\)
gives Eq.~\eqref{eq:localdist}.

For Eq.~\eqref{eq:robust}, replace the actual tester successively by a
Bell-valid tester, a common-slice tester, a field-factorized tester, a
ledger-clean reference realization, and a realization with bounded
backreaction.  Contractivity of total variation distance under each tester and
the triangle inequality give the sum of errors.  More explicitly, distance to
any fixed set is one-Lipschitz:
\begin{equation}
 \big|\dav(p,\cL_{\rm ER})-\dav(q,\cL_{\rm ER})\big|
 \le \dav(p,q).
\end{equation}
Applying this inequality to the replacement chain gives
Eq.~\eqref{eq:robust}.  The result is conditional on experimentally justified
upper bounds for all five replacements; the endpoint Bell statistic supplies
none of them by itself.

\end{document}